\documentclass[letterpaper,twocolumn,10pt]{article}
\usepackage{usenix}
\usepackage{graphicx}
\usepackage[hyphens]{url}
\usepackage{hyperref}
\usepackage[utf8]{inputenc}
\usepackage{array}
\usepackage[usenames]{xcolor}
\begin{document}

\hyphenation{Jon-beshe-Sabz}
\hyphenation{Ma-Bishomarim}
\hyphenation{Proxi-max}

\newcommand{\todo}[1]{{\color{red} #1}}

\date{}

\title{\Large \bf Detecting Censor Detection}

\author{
{\rm David Fifield\thanks{Authors are listed in alphabetical order.}}\\
University of California, Berkeley
\and
{\rm Lynn Tsai}\\
University of California, Berkeley
\and
{\rm Qi Zhong}\\
University of California, Berkeley
}

\maketitle


\subsection*{Abstract}

Our goal is to empirically discover how censors
react to the introduction of new proxy servers
that can be used to circumvent their information controls.
We examine a specific case,
that of obfuscated Tor bridges,
and conduct experiments designed to discover
how long it takes censors to block them
(if they do block at all).
Through a year's worth of active measurements from
China, Iran, Kazakhstan, and other countries,
we learn when bridges become blocked.
In China we found the most interesting behavior,
including long and varying delays before blocking,
frequent failures during which blocked bridges became reachable,
and an advancement in blocking technique midway through the experiment.
Throughout, we observed surprising behavior by censors,
not in accordance with what we would have predicted,
calling into question our assumptions and
suggesting potential untapped avenues for circumvention.


\section{Introduction}

Those who censor the Internet face a twofold challenge:
not only do they have to block direct access to content,
but they also must block access to proxy servers and other
indirect means of circumventing their direct blocks.
Because circumventors continually establish new proxy servers,
effective censorship is therefore an ongoing task,
requiring regular attention and upkeep.
This aspect of the censorship problem---just how
constrained censors are by limitations of resources,
and how it limits their effectiveness---is not well understood,
though circumvention would be improved
by better knowledge of censors' potential weaknesses.
In this research,
we seek to understand the ongoing behavior of censors
as it relates to the specific question of
the blocking of newly introduced Tor bridges.
We do this through frequent active measurements,
in multiple countries,
that allow us to compute the ``lag'' between
when a bridge is introduced and when it becomes blocked.

We limit our inquiry to what should be an
easy case for the censor:
the ``default'' bridges built into Tor Browser.
These bridges use a traffic obfuscation protocol,
their addresses are public,
fairly static, and
easily available to anyone who downloads the browser.
Our usual assumptions about censors
tell us that these bridges should be quickly blocked,
and yet they remain unblocked almost everywhere
in the world, even in places that are known to censor Tor.
Even the famed Great Firewall of China (GFW),
until recently,
was delayed by days or weeks before blocking new bridges.

We ran measurements probes for a year,
testing the reachability of default Tor bridges
every 20 minutes from the U.S., China, Iran, and Kazakhstan.
We recently began collaboration with
established censorship measurement platforms to
expand the tests to more countries.
We found blocking of default bridges
in China and Kazakhstan.
In China, our measurements detected a change in behavior:
around October 2016 the censor switched
from blocking bridges only after release
(after a delay of up to 35 days),
to blocking bridges preemptively.
The blocking in Kazakhstan is qualitatively different
than the blocking in China,
requiring different techniques to detect.


Our results demonstrate that
a discrepancy exists between what we
circumvention researchers assume about censors
and what censors do in practice.
In this work, we take only a few small steps
toward explaining the discrepancy,
performing targeted experiments to learn about
how the censors in China extract bridge addresses
from a software artifact.
We hope to call attention
to potential blind spots and weaknesses of censors
that may be exploited for more effective circumvention.
This is not a call for circumvention researchers
to weaken their threat models;
rather, we hope for richer and more precise threat models
that take into account
underappreciated vulnerabilities
that may lead to more effective circumvention.


\section{Related Work}

There is not much research aimed at systematically
measuring the reactions of censors to the advent of
new or expanded forms of circumvention.
We will list the works we are aware of
that most closely resemble our goals
and elaborate by contrast how our goals are different.

Dingledine~\cite{five-ways-test-bridge-reachability} in 2011
enumerated ways of testing bridge reachability,
among them our primary tool, direct scans.
We attempt to shed light on some of the research questions he laid out,
including knowing what bridges are blocked and where,
and how quickly bridges become blocked.
Dingledine considered the possibility that the very act of testing
the reachability of a bridge could reveal the bridge's existence
to an alert censor.
This consideration is less acute for us,
because we limit ourselves to testing
bridges whose addresses are already known to the public.

There is prior work on careful
distribution strategies
that seek to prevent a censor from discovering
many secret proxy addresses;
examples are
Proximax~\cite{McCoy2011a},
rBridge~\cite{Wang2013a}, and
Salmon~\cite{Douglas2016a}.
These limit the rate at which a censor can
enumerate proxies and maintain a fraction of users whose proxies
remain unblocked.
We, on the other hand, start from a different assumption---that
the proxy addresses are already public---and
observe the nature of censors' actual blocking reactions.

A rich body of research is devoted to
innovating new protocols for disguising traffic,
in order to make it harder to detect and censor.
This present work relies on the properties of a specific protocol,
obfs4~\cite{obfs4}
(which is discussed further in Section~\ref{sec:obfs4}).
Our purpose, however,
is not directly to advance the state of the art of circumvention practice;
but rather to better understand the interaction
between censor and circumventor as it exists today.

Khattak et~al.~\cite{Khattak2013a} showed that understanding censors' models is beneficial to
facilitating evasion. By observing how the GFW processes certain
packets, they were able to deduce some of the underlying weaknesses of the GFW and
suggest ways to exploit them.
Their work is similar to ours in spirit.
While they study ways for a circumventor to defeat on-line detection,
and we study how censors respond to the actions of circumventors,
we have in common the desire for
empirical measurement of actual censor capabilities,

The Open Observatory of Network Interference (OONI),
a censorship measurement platform,
began testing the reachability of
Tor bridges~\cite{ooni-bridge-reachability-study-and-hackfest},
both by simple TCP probes and by
attempts to bootstrap a complete Tor connection.
The tests ran in a few specialized test locations
from March 2014 until February 2015, then mostly lay dormant.
The OONI tests did not specifically examine censors'
treatment of newly introduced bridges.
They revived the test in December 2016 and began to
test the same destinations we were testing.
The results of working with the data of OONI
and another measurement platform, ICLab,
will appear in Section~\ref{sec:ooni-iclab}.

In 2013, Zhu et~al.~\cite{Zhu2013a}
looked into Chinese microblogging sites that implement internal
censorship in order to follow the instructions of Chinese government
restrictions;
specifically, how long it takes for posts to be deleted.
They discuss methods that
detect censorship events within a few minutes of its occurrence. They
discovered that deletions on microblogging sites occurred most frequently
during the initial hour that it was posted, and 30\% occurred in the first 30
minutes. They speculate that the censorship system they use contains a list of
keywords that trigger different censorship behavior, and that if a post is
deleted, most repost chains are deleted within five minutes of the original
post's.
Their work is similar to ours because they, too,
are concerned with time delays.
However, they look at microblog deletion
while we look at proxy blocking.

Nobori and Shinjo~\cite{Nobori2014a} describe the experience of deploying a
circumvention system, VPN Gate, and the Great Firewall's reaction to it, over a
period of two months in 2014.  The GFW blocked VPN Gate's centralized directory
server only three days after initial deployment, and just one day later began
to automatically harvest and block a list of mirror servers.  The next day, the
operators of VPN Gate discovered the IP address of the GFW's automated scanner
and blocked it; in response the GFW began scanning from multiple locations and
cloud services.  VPN Gate began poisoning its list of servers by mixing in
unrelated IP addresses, but after another six days the GFW began verifying
servers as belonging to VPN Gate before blocking them.  During this whole
process, the GFW suddenly and without explanation ceased blocking VPN Gate
servers for about four days, then resumed again.  The lesson of VPN Gate is
that the GFW, at least, is capable of reacting quickly to a new circumvention
system and build automation to block it.

In 2015, Ensafi et~al.~\cite{Ensafi2015b} did a detailed study on how the
GFW uses active probing to quickly and dynamically discover a variety
of types of proxy server.
When the firewall detected a suspicious
connection, it would send a followup request to the destination address to
identify it. They noted that this type of probing for Tor
already existed in early 2013.
The discovery of active probing
led to the development of probing-resistant protocols,
obfs4 among them.
That a national censor can be so sophisticated in some ways
(active probing)
and yet seem to lag in others (blocking of default bridges)
is a challenge to our mental models.

\section{Background}

Tor~\cite{tor} is an anonymity network
that is also used for censorship circumvention.
Tor's resistance to censorship is due not to
anything inherent in the protocol itself;
but to its surrounding infrastructure
of \emph{bridges} and \emph{pluggable transports}.
Bridges~\cite{Dingledine2006a} are secret Tor servers,
the addresses of which are not widely distributed,
preventing easy discovery by a censor.
Users are meant to learn a few bridge addresses
through an out-of-band channel, like email.
Pluggable transports~\cite{pluggable-transports} are covert communications
protocols that disguise Tor's traffic signature
on the wire, preventing easy online detection.
Bridges that use a particular pluggable transport,
called obfs4, are the main focus of our study.
The properties of obfs4 are covered in Section~\ref{sec:obfs4}.

Tor Browser~\cite{torbrowser} is the means by which
most ordinary users access the Tor network,
whether for anonymity or censorship circumvention.
It is a modified version of Firefox
with a built-in Tor client
and a special interface
for the configuration of bridges and pluggable transports.

Separate from the infrastructure of secret bridges,
Tor Browser also ships with a number of
built-in, \emph{default} bridges,
whose addresses
are baked into the source code~\cite{torbrowser-bridgeprefs}.
There is a configuration file inside the Tor Browser
listing all these default bridges. 
Users can use a default bridge simply by selecting a
pluggable transport from a menu,
no out-of-band communication required.
Strictly speaking, a ``default bridge''
is a contradiction:
bridges are supposed to be secret,
not easily discoverable in a configuration file.
Our intuition and the common assumptions
in censorship research tell us that the
default bridges, which are trivially discoverable,
should be quickly blocked---and yet they are not.
Indeed, a study by Matic et.~al~\cite{Matic2017a} found that
over 90\% of bridge users use one of the default bridges.
Even the uncommonly capable Great Firewall of China,
before October 2016, delayed for days or weeks before
blocking these bridges,
in contrast to the rapidity with which
it blocks other, harder-to-detect proxies.
The speed with which censors block default Tor Browser bridges
is the main object of our study.

Besides Tor Browser,
there is also an Android version of Tor called Orbot.
Orbot and Tor Browser have most of their default bridges in common,
but a few appear in Orbot only.
We will be looking at these set of different bridges as well.

\subsection{Tor Browser releases}
Tor Browser releases changes in two different tracks: stable and alpha.
The stable track is ``safe'' so to speak, 
with only small changes at any given time.
For the most part, stable releases include bug fixes.
On the other hand, the alpha track is much more experimental.
It contains experimental features that are in ``test'' until it matures
and can be merged with the stable track.
Stable and alpha releases tend to appear around the same time since 
both of them are driven by Firefox releases.
If a Tor Browser version number contains the letter `a',
it is an alpha version (for example, 6.5a3).
Upon a new Firefox release, Tor Browser, too, updates any security
flaws that may have been discovered.
Each release is an opportunity for Tor Browser to release new bridges.
During our study over a duration of approximately 12 months, 
we observed a total of 18 stable releases and 13 alpha releases~\cite{torbrowser-changelog}.

\subsection{The lifecycle of a new bridge}
In the following sections, we elaborate on the lifecycle of a new bridge 
and the stages involved in releasing it.
Each stage in the lifecycle is also a potential opportunity for censors 
to detect the addition of a new bridge.

\begin{enumerate}
    \item \textbf{Ticket Filed:} Filing a ticket in Tor's online bug tracker
        proposes the inclusion of new default bridges. Censors may monitor the
        bug tracker and discover the new default bridges here.
    \item \textbf{Ticket Merged:} Bridges are added to the Tor Browser's source
        code during the merging of a ticket. It is automatically included in
        nightly builds, and code containing the new bridge is available in
        executable form. Censors may learn of bridges at this stage if they are
        looking at the source code repository or monitoring nightly builds.
    \item \textbf{Testing Release:} Preceding a public release, Tor Browser
        developers prepare candidate packages and send them out to the quality
        assurance mailing list for testing. Censors monitoring the mailing list
        could discover new bridges at this point. 
    \item \textbf{Public Release:} After bridges have been tested, the new
        packages are announced on the Tor Blog. The Tor Blog is publicly
        available, and any censors monitoring it would discover the new bridges
        here. Installed Tor Browsers will automatically update to include the
        new packages, and users will begin to actively use the new bridges. A
        censor could also discover the new bridges through black-box testing an
        auto-updating installing at this stage. 
\end{enumerate}

The entirety of the lifecycle usually takes a few weeks to complete.
Occasionally, if the fix is small enough, the Testing Release stage is skipped.
There is also a possibility that new bridges are discussed 
in private mailing lists beforehand, 
and a censor that had infiltrated the mailing lists could 
conceivably discover new bridges before a ticket had even been filed.

\subsection{The Properties of obfs4}
\label{sec:obfs4}
In our studies, we focus primarily on a particular pluggable transport: obfs4,
an advanced transport protocol that offers
resistance to deep packet inspection
and resistance to active probing.

\begin{itemize}
    \item \textbf{Deep Packet Inspection} Re-encrypts a Tor stream so that it
        appears as a stream of random bytes that cannot be easily decrypted. 
    \item \textbf{Active Probing Attacks} Censors scan suspected proxies in
        order to discover what protocols are supported.  The Great Firewall is
        known to use active proving against predecessor protocols obfs2 and
        obfs3. However, this attack does not work with obfs4. Every obfs4
        bridge has a per-bridge secret in which a client must prove knowledge
        of upon the initial message. The censor would have to shave the same
        out-of-band information as a legitimate client, therefore rendering the
        knowledge of a bridge IP address insufficient to prove the existence of
        a bridge.
\end{itemize}

These security features are crucial because they allow us to limit the methods
of bridge discovery. 
They give us the confidence that censors discover our bridges in the ways that
we intend them to.

obfs4 is an important bridge not only for its unique properties,
but also for its applications in the real world: 
it is the most commonly used bridge, 
yielding about 35,000 concurrent users in February 2017~\cite{obfs4-users}.

\section{Methodology}

\begin{figure*}
\includegraphics{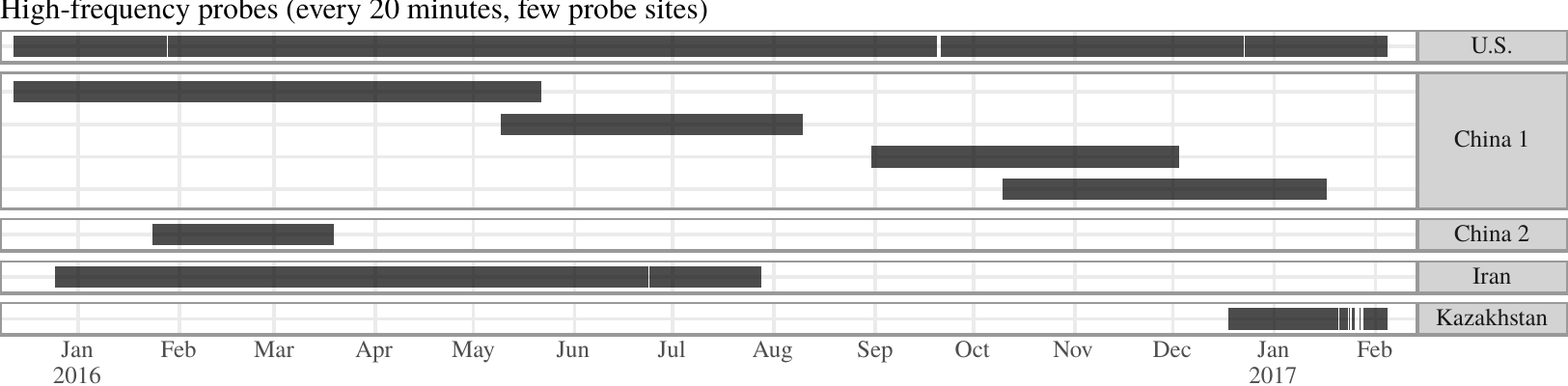}

\vspace{1mm}

\includegraphics{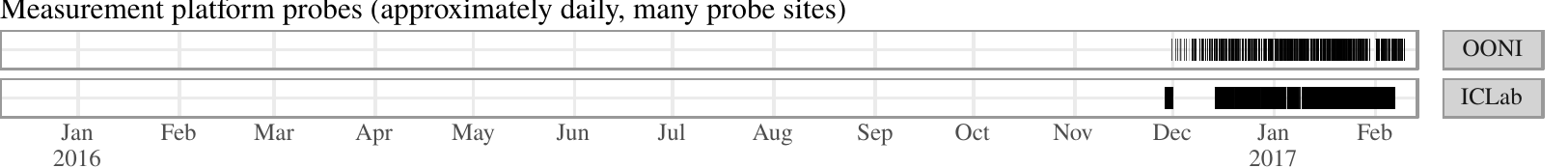}
\caption{
Time coverage of our
active measurements of bridge reachability.
In the upper portion of the figure, ``High-frequency probes,''
each box represents one autonomous system
and each row is a distinct IP address.
We had four probe sites in the ``China~1'' autonomous
partially overlapping in time,
which we spliced together into one series.
The measurement platforms OONI and ICLab,
in the lower portion of the figure,
cover many more countries but run less frequently.
}
\label{fig:timespans}
\end{figure*}

Our experiment requires us to be able to detect
the moment a bridge is blocked:
when the bridge transitions from being reachable
to being unreachable.
We do this primarily through active measurements
from probe sites located in various countries.

For a little more than a year,
we ran frequent TCP reachability tests
of a variety of destinations
from probe sites in the U.S., China, Iran, and Kazakhstan.
Because of the difficulty of acquiring test machines
inside countries subject to information controls,
probe coverage is not continuous or complete in any
country other than the U.S.
The probe site in the U.S. acted as a control,
allowing us to distinguish occurrences of blocking from temporary bridge outages.
Figure~\ref{fig:timespans} shows the time periods
for which we have measurements in each country.
From each probe site, we attempted a TCP connection
to every destination
every 20 minutes,
recording for each connection attempt
whether the connection was successful,
the time elapsed,
and error message if any.
The rate of probing enables us to know not only the
date, but also the time of day,
when each bridge became blocked.
The set of destinations,
a mix of fresh default obfs4 bridges and other bridges,
appears in Table~\ref{tab:destinations}.

\begin{table}
\small
\begin{center}
\begin{tabular}{@{\quad}r@{~:~}l}
nickname & ports \\
\hline
\noalign{\smallskip}
\multicolumn{2}{l}{\textbf{New Tor Browser default obfs4 bridges}} \\
ndnop3 & 24215, 10527 \\
ndnop5 & 13764 \\
riemann & 443 \\
noether & 443 \\
Mosaddegh & 41835, 80, 443, 2934, 9332, 15937 \\
MaBishomarim & 49868, 80, 443, 2413, 7920, 16488 \\
GreenBelt & 60873, 80, 443, 5881, 7013, 12166 \\
JonbesheSabz & 80, 1894, 4148, 4304 \\
Azadi & 443, 4319, 6041, 16815 \\
Lisbeth & 443 \\
NX01 & 443 \\
LeifEricson & 50000, 50001, 50002 \\
\noalign{\smallskip}
\multicolumn{2}{l}{\textbf{Already existing Tor Browser default bridges}} \\
LeifEricson & 41213 \\
fdctorbridge01 & 80 \\
\noalign{\smallskip}
\multicolumn{2}{l}{\textbf{Orbot-only default obfs4 bridges}} \\
Mosaddegh & 1984 \\
MaBishomarim & 1984 \\
JonbesheSabz & 1984 \\
Azadi & 1984 \\
\noalign{\smallskip}
\multicolumn{2}{l}{\textbf{Never-published obfs4 bridges}} \\
ndnop4 & 27668 \\
\end{tabular}
\end{center}
\caption{
The destinations we tested,
consisting mostly of new obfs4 bridges,
along with some old or never-published bridges.
A bridge is identified by its ``nickname,''
an arbitrary label chosen by its operator.
Each nickname represents an IP address.
Multiple ports on the same IP address
count as distinct bridges for our purposes.
We also tested
port 22 (SSH) on the bridges that had it open.
}
\label{tab:destinations}
\end{table}

During the latter part of our measurements,
we got assistance from the established
censorship measurements platforms OONI and ICLab.
Razaghpanah et~al.~\cite{RazaghpanahLFNV16}
describe both platforms, their similarities
and different design tradeoffs.
At our request, both platforms added to their
repertoire of measurements
active measurement of default Tor Browser bridges.
Compared to our custom reachability tests,
the ICLab and OONI measurements trade
frequency for coverage:
they run daily, rather than every 20 minutes,
but they cover many more geographic locations,
giving a more global view of censorship.
Our OONI-derived data covers 117 ASes in 55 countries
and our ICLab-derived data covers 201 ASes in 217 countries.
(The actual division is by two-letter country code,
of which there are more than there are countries in the world.
ICLab heavily relies on measurements from VPN endpoints,
including some in autonomous systems that span more than one country.)


Simple TCP reachability testing has limitations,
in that a censor may make a bridge effectively useless,
without directly blocking its IP address or TCP port.
China blocks at the TCP/IP layer,
so block are easy to detect.
On the other hand,
we found that Kazakhstan blocks at a higher layer.
A successful TCP connection doesn't necessarily mean a successful Tor connection.
For this reason, for a limited time we also
did testing of full Tor-over-obfs4 connections.
Details of this experiment appear in
Section~\ref{sec:limitations}.

Throughout the measurements,
we coordinated with the developers of Tor Browser
to begin measurement of bridges before their introduction.
During the course of the study,
the Tor Project was ramping up its obfs4 capacity
by adding additional bridges.
We additionally ran certain controlled experiments
designed to uncover specific blocking behaviors of the censor.
These included changing ports on the same address,
and inserting a bridge
so that it is available in the same code
but commented out.

\section{Results and Observations}


\begin{figure*}
\includegraphics{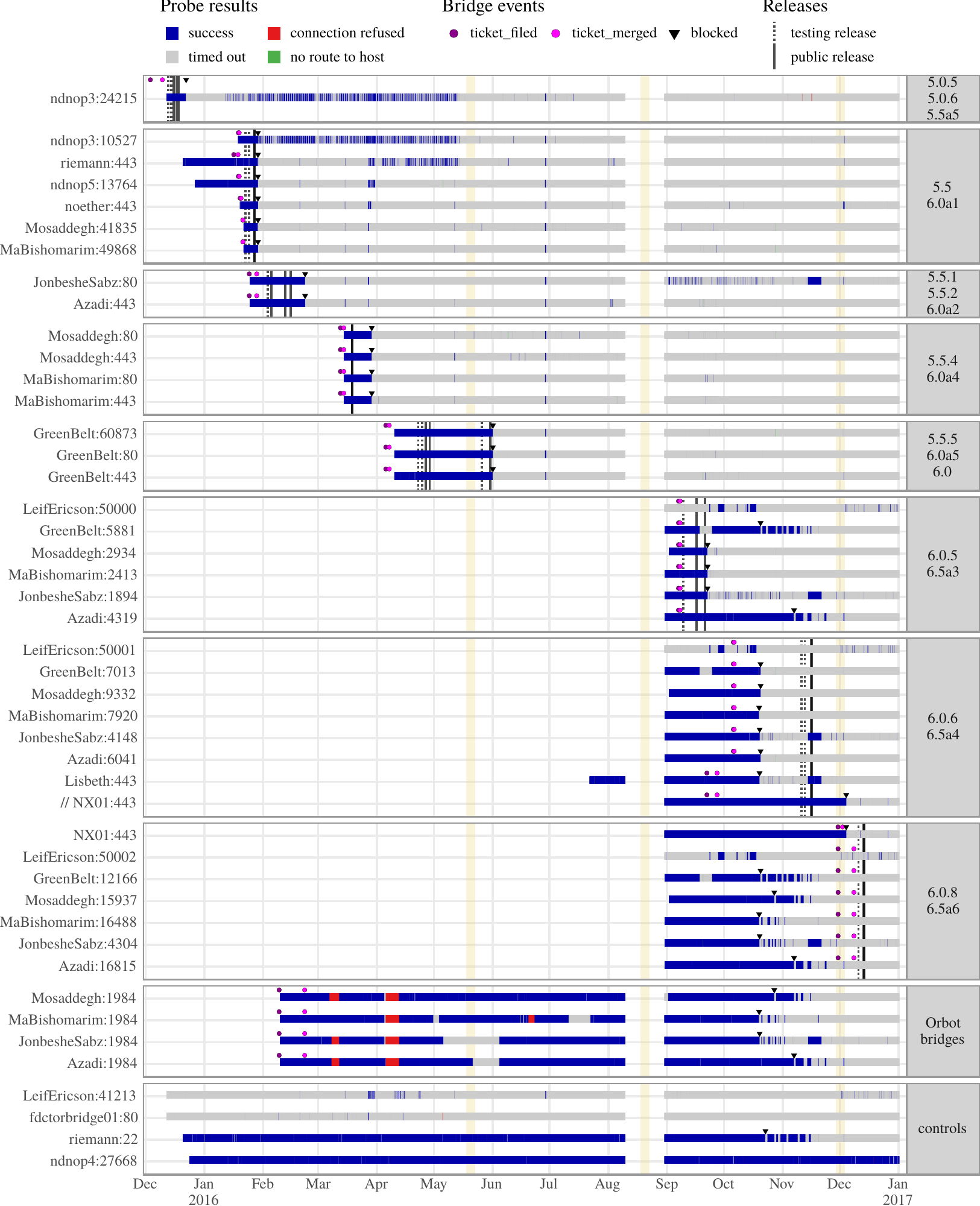}
\caption{
Timeline of Tor
Browser default bridge reachability in the ``China~1'' AS.
Black vertical lines (dashed and solid) indicate releases.
There is a gap in the data between August~9 and August~30, 2016.
Before October~2016, bridges were
blocked only after a release; after that they began to be blocked as
soon as their ticket was merged.  The vertical yellow stripes indicate
where observations from different sites were spliced together.
The
notation ``//~NX01:443'' indicates that the bridge was commented out in
this release, and uncommented in the following release.
}
\label{fig:timelines}
\end{figure*}

\newcommand{\eventheading}[1]{\textbf{#1}\quad}
\newcommand{\ticketfiled}{\eventheading{Ticket filed}}
\newcommand{\ticketmerged}{\eventheading{Ticket merged}}
\newcommand{\testrelease}{\eventheading{Testing release}}
\newcommand{\pubrelease}{\eventheading{Public release}}
\newcommand{\blocked}{\eventheading{Blocked}}
\newcommand{\portrunning}{\eventheading{Ports Running}}
\newcommand{\releaseheading}[2]{\noalign{\smallskip\normalsize\textbf{#1:} #2}}
\newcommand{\releaseheadingnull}[1]{\noalign{\smallskip\normalsize\textcolor{gray}{\textbf{#1:} no new bridges}}}
\begin{table*}
\footnotesize
\setlength{\tabcolsep}{2pt}
\begin{minipage}[t]{3.375in}
\vspace{0pt}
\begin{tabular}{l r >{\raggedright\arraybackslash}p{2.2in}}

\releaseheading{Tor Browser 5.0.5/5.0.6/5.5a5}{1 new bridge}
03 Dec & $-$19~days & \ticketfiled ndnop3:24215 \\
09 Dec & $-$13~days & \ticketmerged ndnop3:24215 \\
12 Dec &  $-$9~days & \testrelease 5.0.5 stable \\
14 Dec &  $-$8~days & \testrelease 5.5a5 alpha \\
15 Dec &  $-$7~days & \pubrelease 5.0.5 stable \\
17 Dec &  $-$5~days & \pubrelease 5.0.6 stable \\
18 Dec &  $-$4~days & \pubrelease 5.5a5 alpha \\
22 Dec &     0\phantom{~days} & \blocked ndnop3:24215 \\

\releaseheadingnull{Tor Browser 5.0.7/5.5a6}

\releaseheading{Tor Browser 5.5/6.0a1}{6 new bridges}
16 Jan & $-$13~days & \ticketfiled riemann:443 \\
18 Jan & $-$10~days & \ticketmerged riemann:443 \\
18 Jan & $-$10~days & \ticketfiled ndnop3:10527, ndnop5:13764 \\
19 Jan & $-$10~days & \ticketmerged ndnop3:10527, ndnop5:13764 \\
19 Jan &  $-$9~days & \ticketfiled noether:443 \\
20 Jan &  $-$9~days & \ticketmerged noether:443 \\
21 Jan &  $-$8~days & \ticketfiled Mosaddegh:41835, MaBishomarim:49868 \\
21 Jan &  $-$8~days & \ticketmerged Mosaddegh:41835, MaBishomarim:49868 \\
22 Jan &  $-$7~days & \testrelease 5.5 stable \\
24 Jan &  $-$5~days & \testrelease 6.0a1 alpha \\
27 Jan &  $-$2~days & \pubrelease 5.5 stable \\
27 Jan &  $-$2~days & \pubrelease 6.0a1 alpha \\
29 Jan &     0\phantom{~days} & \blocked ndnop3:10527, ndnop5:13764, noether:443, riemann:443, MaBishomarim:49868, Mosaddegh:41835 \\

\releaseheading{Tor Browser 5.5.1/5.5.2/6.0a2}{2 new bridges}
24 Jan & $-$29~days & \ticketfiled JonbesheSabz:80, Azadi:443 \\
28 Jan & $-$26~days & \ticketmerged JonbesheSabz:80, Azadi:443 \\
03 Feb & $-$20~days & \testrelease 5.5.1 stable \\
05 Feb & $-$18~days & \pubrelease 5.5.1 stable \\
12 Feb & $-$11~days & \pubrelease 5.5.2 stable \\
15 Feb &  $-$8~days & \pubrelease 6.0a2 alpha \\
23 Feb &     0\phantom{~days} & \blocked JonbesheSabz:80, Azadi:443 \\

\releaseheadingnull{Tor Browser 5.5.3/6.0a3}

\releaseheading{Tor Browser 5.5.4/6.0a4}{4 new bridges}
12 Mar & $-$16~days & \ticketfiled Mosaddegh:80, Mosaddegh:443, MaBishomarim:80, MaBishomarim:443 \\
14 Mar & $-$15~days & \ticketmerged Mosaddegh:80, Mosaddegh:443, MaBishomarim:80, MaBishomarim:443 \\
18 Mar & $-$10~days & \pubrelease 5.5.4 stable \\
18 Mar & $-$10~days & \pubrelease 6.0a4 alpha \\
29 Mar &     0\phantom{~days} & \blocked Mosaddegh:443, Mosaddegh:80, MaBishomarim:443, MaBishomarim:80 \\

\releaseheading{Tor Browser 5.5.5/6.0a5/6.0}{3 new bridges}
05 Apr & $-$56~days & \ticketfiled GreenBelt:60873, GreenBelt:443, GreenBelt:80 \\
07 Apr & $-$55~days & \ticketmerged GreenBelt:60873, GreenBelt:80, GreenBelt:443 \\
22 Apr & $-$39~days & \testrelease 5.5.5 stable \\
24 Apr & $-$37~days & \testrelease 6.0a5 alpha \\
26 Apr & $-$35~days & \pubrelease 5.5.5 stable \\
28 Apr & $-$33~days & \pubrelease 6.0a5 alpha \\
26 May &  $-$6~days & \testrelease 6.0 stable \\
30 May &  $-$1\rlap{~day}\phantom{~days} & \pubrelease 6.0 stable \\
01 Jun &     0\phantom{~days} & \blocked GreenBelt:60873, GreenBelt:443, GreenBelt:80 \\

\end{tabular}
\end{minipage}%
\begin{minipage}[t]{3.375in}
\vspace{0pt}
\begin{tabular}{l r >{\raggedright\arraybackslash}p{2.2in}}

\releaseheadingnull{Tor Browser 6.0.1/6.0.2/6.0.3/6.0.4/6.5a1/6.5a2}

\releaseheading{Tor Browser 6.0.5/6.5a3}{6 new bridges}
?? & ??\phantom{~days} & \blocked LeifEricson:50000 \\
06 Sep & $-$15~days & \ticketfiled LeifEricson:50000, GreenBelt:5881, Mosaddegh:2934, MaBishomarim:2413, JonbesheSabz:1894, Azadi:4319 \\
07 Sep & $-$14~days & \ticketmerged LeifEricson:50000, GreenBelt:5881, Mosaddegh:2934, MaBishomarim:2413, JonbesheSabz:1894, Azadi:4319 \\
09 Sep & $-$13~days & \testrelease 6.0.5 stable \\
16 Sep &  $-$6~days & \pubrelease 6.0.5 stable \\
20 Sep &  $-$1\rlap{~day}\phantom{~days} & \pubrelease 6.5a3 alpha \\
22 Sep &     0\phantom{~days} & \blocked Mosaddegh:2934, MaBishomarim:2413, JonbesheSabz:1894 \\
20 Oct & $+$28~days & \blocked GreenBelt:5881 \\
06 Nov & $+$46~days & \blocked Azadi:4319 \\

\releaseheading{Tor Browser 6.0.6/6.5a4}{8 new bridges}
?? & ??\phantom{~days} & \blocked LeifEricson:50001 \\
21 Sep & $-$28~days & \ticketfiled Lisbeth:443, // NX01:443 \\
27 Sep & $-$22~days & \ticketmerged Lisbeth:443, // NX01:443 \\
05 Oct & $-$14~days & \ticketfiled LeifEricson:50001, GreenBelt:7013, Mosaddegh:9332, MaBishomarim:7920, JonbesheSabz:4148, Azadi:6041 \\
06 Oct & $-$13~days & \ticketmerged LeifEricson:50001, GreenBelt:7013, Mosaddegh:9332, MaBishomarim:7920, JonbesheSabz:4148, Azadi:6041 \\
19 Oct &     0\phantom{~days} & \blocked MaBishomarim:7920, Lisbeth:443, JonbesheSabz:4148 \\
20 Oct &  $+$1\rlap{~day}\phantom{~days} & \blocked GreenBelt:7013, Azadi:6041, Mosaddegh:9332 \\
10 Nov & $+$22~days & \testrelease 6.0.6 stable \\
12 Nov & $+$24~days & \testrelease 6.5a4 alpha \\
15 Nov & $+$27~days & \pubrelease 6.0.6 stable \\
16 Nov & $+$28~days & \pubrelease 6.5a4 alpha \\
04 Dec & $+$46~days & \blocked // NX01:443 \\

\releaseheadingnull{Tor Browser 6.0.7/6.5a5}

\releaseheading{Tor Browser 6.0.8/6.5a6}{7 new bridges}
?? & ??\phantom{~days} & \blocked LeifEricson:50002 \\
19 Oct &     0\phantom{~days} & \blocked MaBishomarim:16488, JonbesheSabz:4304 \\
20 Oct &  $+$1\rlap{~day}\phantom{~days} & \blocked GreenBelt:12166 \\
27 Oct &  $+$8~days & \blocked Mosaddegh:15937 \\
06 Nov & $+$18~days & \blocked Azadi:16815 \\
30 Nov & $+$42~days & \ticketfiled NX01:443, LeifEricson:50002, GreenBelt:12166, Mosaddegh:15937, MaBishomarim:16488, JonbesheSabz:4304, Azadi:16815 \\
02 Dec & $+$44~days & \ticketmerged NX01:443 \\
04 Dec & $+$46~days & \blocked NX01:443 \\
08 Dec & $+$50~days & \ticketmerged LeifEricson:50002, GreenBelt:12166, Mosaddegh:15937, MaBishomarim:16488, JonbesheSabz:4304, Azadi:16815 \\
10 Dec & $+$52~days & \testrelease 6.0.8 stable \\
13 Dec & $+$55~days & \pubrelease 6.0.8 stable \\
13 Dec & $+$55~days & \pubrelease 6.5a6 alpha \\

\end{tabular}
\end{minipage}%

\caption{
Blocking of bridges in the ``China~1'' data set.
Releases are batched according to the fresh bridges they contain.
Time offsets are relative to the ``0'' date of first blocking
within each batch.
The notation ``//~NX01:443'' indicates a commented-out bridge.
Before the release batch 6.0.6/6.5a4, bridges are blocked only after a release;
after that they are blocked preemptively.
Notice the concurrent blocking dates in the 6.0.6/6.5a4 and 6.0.8/6.5a6 batches.
}
\label{table:timeline}
\end{table*}


In this section we present and interpret the results
of our experiments, focusing on China,
where we had the most measurements
and saw the most varied behavior.
Our observations in the ``China~1'' and ``China~2'' ASes
were mostly in agreement;
minor differences are mentioned in Sections~\ref{sec:anomalies} and~\ref{sec:IPBlock}.
In Iran, we did not see any blocking of bridges;
all of them were reachable all the time---though
what we discovered in Kazakhstan means that
there may have been blocking that TCP reachability tests would not detect.
We found blocking of the default bridges in Kazakhstan,
though of a qualitatively different nature
than that which we observed in China.
We cover the particulars of Kazakhstan in Section~\ref{sec:limitations}.
Throughout this section, refer to
Figure~\ref{fig:timelines} and Table~\ref{table:timeline},
which depict the entirety of the combined ``China~1'' data set.

Overall, we recorded over 5.9~million individual probe results.
Our high-frequency probes account for 4.9~million of these;
ICLab accounts for about 800,000 and OONI for about 260,000.
There are 2.1~million probe results in the ``China~1'' AS alone,
which is the basis for Figure~\ref{fig:timelines}.

We have organized Tor Browser releases into ``batches'',
where each batch contains a distinct set of fresh bridges.
Figure~\ref{fig:timelines} and Table~\ref{table:timeline}
are arranged by release batch.
During the first part of our experiment,
blocking events were distinct:
when a batch contained more than one bridge,
all were blocked at once (within our probing period of 20 minutes).
In our first six batches, we observed blocking delays of
7, 2, 18, 10, 35, and~6 days after the first public release,
and up to 57 days
after the filing of the first ticket,
when bridges were potentially first discoverable.
The only exception to this was that in the 6.0.5/6.5a3 batch,
the censor actually failed to blocked two bridges,
and these two bridges were blocked only much later.
This fact suggests, to us, that new default bridges
are loaded into the firewall in groups,
and are not, for example, detected and blocked one at a time.
During the first six batches,
we found that blocking in China was keyed on both IP address and port,
consistent with an observation of Winter and Lindskog in 2012~\cite{Winter2012a}.
For example, many of the bridges happened to have port 22 (SSH) open,
and it remained accessible even as other ports on the same IP address were blocked.
(See riemann in Figure~\ref{fig:timelines} for an example:
its port 22 remained accessible when its port 443 was blocked in January 2016.)
Per-port blocking is what enabled us to run multiple bridges
on the same IP address.

In the last two batches,
we noticed that the GFW seems to have altered their blocking methods.
New bridges in these two batches were all blocked even before the public release.
They were blocked soon after they were merged into the public Git repository.

During this period, we observed that China also started blocking on whole IP address,
as well as continuing blocking on IP address and port pairs.
Unlike blocking bridges, block for whole IP addresses did not seem to be done all at once.
By the end of December 2016,
all our bridges were blocked on the whole IP.
Running any more bridges on existing IPs is no longer possible.

As for Orbot bridges, we found that China did not try to block them at all.
They remained accessible until late October and early November in 2016.
At this point, China started blocking whole IP address of bridges in Tor Browser.
Since these Orbot bridges only used different ports on the same IPs,
they were blocked as a side affect.

\subsection{Port Rotations}

In our first few batches, we found that China blocked on IP and port pairs.
This means that when we have an existing bridge that is blocked,
if we just open up a new port on the same IP,
the new port would still be reachable.
This is an easy way of evading GFW blocks
compared to setting up new bridges.
We call this process port rotation.
We were interested in seeing whether or not
rotating the ports would give us new results.
In release batches 6.0.5/6.5a3, 6.0.6/6.5a4, and 6.0.8/6.5a6,
we changed the port number of certain existing bridges
with each new batch,
creating what appeared to the GFW to be a large set of new bridges each time.
Table~\ref{tab:portrotation} displays the ports
that we rotated to.

\begin{table}
\setlength{\tabcolsep}{0pt}
\begin{center}
\begin{tabular}{r@{~:~}r@{$\rightarrow$}r@{$\rightarrow$}r@{$\rightarrow$}r}
LeifEricson & 41213 & 50000 & 50001 & 50002 \\
GreenBelt & 60873 & 5881 & 7013 & 12166 \\
Mosaddegh & 41835 & 2934 & 9332 & 15937 \\
MaBishomarim & 49868 & 2413 & 7920 & 16488 \\
\multicolumn{3}{r}{JonbesheSabz~:~1894$\rightarrow$} & 4148 & 4304 \\
\multicolumn{3}{r}{Azadi~:~4319$\rightarrow$} & 6041 & 16815 \\
\end{tabular}
\caption{
Rotation of port numbers in successive releases.
The strategy worked until the second-to-last time,
when the GFW began blocking entire IP addresses.
The ports in the final rotation were blocked
even before they were used.
}
\label{tab:portrotation}
\end{center}
\end{table}

The first rotation was successful.
The new ports served as new unblocked bridges,
and worked for a time after release before being blocked,
as before.
The second rotation was initially successful,
but this time when the bridges were blocked,
all ports on the IP address were blocked,
including the ports we had reserved for the third rotation.
The status of the bridges after release 6.0.5 is shown in Table~\ref{table:blocked}.  

\begin{table}
\begin{center}
\begin{tabular}{ll}
\textbf{Bridge} & \textbf{Status} \\
\hline
LeifEricson:50000  & Blocked \\
GreenBelt:5881 & Unblocked$^*$ \\
Mosaddegh:2934 & Blocked \\
MaBishomarim:2413 & Blocked \\
JonbesheSabz:1894 & Blocked \\
Azadi:4319 & Unblocked \\
\end{tabular}
\caption{
Blockage status in the days following the 6.0.5/6.5a3 releases.
LeifEricson had been blocked since we started measuring it.
GreenBelt had an outage on all ports during the time of blocking,
which may have protected it when the other bridges were blocked.
It was not blocked when it recovered from its outage.
and was also unblocked once it recovered from its outage.
Azadi:4319 somehow eluded discovery when the other bridges were blocked,
and remained unblocked for a while even after Azadi:6041 was blocked in the following release.
}
\label{table:blocked}
\end{center}
\end{table}

\subsection{Failure to Block All New Bridges in a Single Release}
\label{sec:differenttimes}

In batch 6.0.5/6.5a3, we rotated six ports, one (LeifEricson) was preemptively blocked on
all ports, three were blocked on the same day, and two (GreenBelt and Azadi)
were not blocked. This is the first time we have seen this phenomenon
over the past year of observation. One of the unblocked one (GreenBelt) was not
operational at the time, but did come back online later.
We will now look at each of
these two bridges in detail.

\subsubsection{GreenBelt}

Our data shows that the U.S. probe
site was not able to connect to GreenBelt for an extended period of time,
namely from September~17, 2016 to September~24, 2016.
After inquiring with the bridge operator,
we found that GreenBelt was indeed down during this time due to an IP table
configuration error. The bridge rejected any incoming traffic. The blocking of
bridges for this release happened on September~22, 2016, which means that GreenBelt
was not functioning when the blocking happened. This is a strong indicator that
the censor used network analysis techniques rather than parsing the bridge
configuration file directly.

\subsubsection{Azadi}

Both our observation data and the bridge operator
confirmed that Azadi had been working properly, unlike GreenBelt. Combined with
the fact that this is the only time it happened during our observation,
it shows that the censor's method for finding new
bridges would have a low probability of missing new bridges. This confirms our
previous suspicion that the censor is not parsing the bridge configuration
file. 
%
%
%

\subsubsection{Analysis}

Our speculation for this anomaly is that the censor
used black-box network traffic analysis to find new bridges. In other words,
they ran the released version of Tor and simply monitored what addresses the
executable connected to. Since GreenBelt was down, this method would have
missed it. GreenBelt being down at the time could provide a reason for it not
getting blocked, were it not for the fact that Azadi did not have an outage and
it also did not get blocked.
Tor Browser at that point had a very large number of default bridges included.
If the censor did not monitor the executable long enough,
this type of traffic analysis
might miss some new bridges. We believe this is what happened to Azadi.

\subsection{Preemptive Blocking}

During our studies,
it appears as though there was a change in the GFW's method
of bridge discovery in October 2016.
Rather than wait until after a release to block bridges,
it started blocking them after a ticket was merged (before release).
In release 6.0.6/6.5a4, we can see in Table \ref{table:timeline}
that all the new bridges were blocked before the release.
This behavior is drastically different from what we have seen before.
Assuming that the censor has not infiltrated the private mailing list,
Tor bridges appear in two places before a release.
When requesting to add new bridges,
a ticket has to be submitted to the Tor bug tracker.
When this ticket is accepted,
it is merged and the new bridges would be added to the source code.
Both the Tor bug tracker and the source code repository are publicly accessible,
so a censor could learn the new bridges from either of these two places.
We noted though that learning new bridges from the tickets
would require human inspection,
since the tickets do not have specific formats and can be on any issue.
On the other hand,
a script can easily keep track of new bridges from the source code repository.
It would be easier for them to read from the source code repository.

\subsection{Batch Blocking Timing patterns}
\label{sec:timingPattern}

From Table~\ref{table:batchBlockTime},
we can see that for our eight releases,
the batch blocking all happened on weekdays.
Most of the blocking happened on Tuesday and Thursday.
Only one blocking happened on Wednesday, and one on Friday.
Furthermore, they all happened between 10:00 and 17:00 local time.
There seem to be no noticeable patterns to these time.
However, they all lie within working hours in China.
This seems to indicate some manual effort is needed
to make blocking take affect.
The idea of manual effort is further supported
by the delay we see in blocking for the first six batches.
If the blocking is done purely automatically,
we would not expect to see a varying of delay between the release and blocking.

One other thing to note is that starting from version 6.0.6/6.5a4,
bridges were blocked before a release.
It suggests that the blocking process might have changed as well,
and the process might be automatic now.

When the GFW blocks whole IP addresses,
they seem to have a different timing pattern than the batch blocking of bridges.
This will be discussed in more detail in \ref{sec:IPBlock}.

\subsection{IP Blocks}
\label{sec:IPBlock}

\begin{table}
\begin{center}
\begin{tabular}{lccl}
\textbf{Bridge} & \textbf{Time} & \textbf{Day} & \textbf{Date}\\
\hline
GreenBelt & 14:00 & Thurs & 20 Oct 2016\\
Mosaddegh & 19:40 & Thurs & 27 Oct 2016\\
MaBishomarim & 20:20 & Wed & 19 Oct 2016\\
JonbesheSabz & 02:00 & Thurs & 20 Oct 2016\\
Azadi & 05:50 & Mon & 07 Nov 2016\\
\end{tabular}
\caption{
IP Blocking Time
in CST time. Since there are interweaving unblocking after the initial IP
block, we are only looking at the initial block time here.
}
\end{center}
\end{table}

In the first six release batches,
we confirmed a finding of Winter and Lindskog~\cite{Winter2012a}
that the Great Firewall blocks bridges by their specific port number.
A bridge that is discovered and blocked on one port will
not cause other ports on the IP address to get blocked.
This property of the firewall allowed us to rotate ports
and, for example, take the MaBishomarim bridge through ports
49868, 80, 2413,
in successive releases.

Since October 2016 and the 6.0.6/6.5a4 release batch, bridges have been
blocked on the entire IP. On
October 20, two bridges were blocked together on the entire IP address.
However, we still could not account for the blocking of
Mosaddegh and Azadi, which happened on a seemingly unrelated date.
Blocks affected even ports not-yet-used ports that were waiting
in reserve for the following release,
and non-bridge ports such as 22.
The all-ports blocking affected even one of the earliest bridges
we had measured, riemann, whose obfs4 port was blocked in January 2016
and its SSH port 10 months later (see Figure~\ref{fig:timelines}).

%
%
%

Between November~15
and November~21, 2016, JonbesheSabz and Lisbeth were reachable from one site
in the ``China~1'' AS, but
not the other.

As mentioned in section~\ref{sec:timingPattern}, all simultaneous blocking for bridges
happened during working hours on a weekday. This suggested manual blocking
instead of an automatic blocking system. One key observation we have is that
blocking for IP occurred at times such as 2:00AM and 5:50AM China Standard
Time, likely falling outside the range of the standard workday. This allows us
to reach a few possible scenarios: 

\begin{enumerate}
\item The limitation to working hours is just a coincidence,
an artifact of the small size of the blocking event data set.
\item Individual port blocking uses a different system
than IP:port blocking.
Even though IP:port
blocking is manual, IP address blocking is automatic.
\end{enumerate}

Another observation is that there is a pattern of interweaving unblocking and
blocking after an IP block.
Although Mosaddegh
was already blocked on October 27, it later became available again for a period
of time whereas previously blocked individual ports remained blocked. It is
only after November 15 that it became permanently blocked. We observe this
pattern across all the blocked bridges. Also, we notice that even when the IP
address was unblocked, ports that are previously blocked would still stay blocked. In
the example above, we can see that port 9332 and 2934 for Mosaddegh were never
unblocked even though the IP itself was unblocked multiple times. This seems to
indicate a two-tier structure to the GFW. The IP:port blocking and the IP
blocking are handled separately. If an IP:port pair is in either of these,
it would get blocked.


One possible hypothesis is that this is an
artifact resulted from GFW could not handle so much traffic. The
interweaving unblocking and blocking could simply result from GFW failing under
too much traffic. Currently we don't really know whether our hypothesis is correct,
but this would be part of our intended future work.

\begin{table}
\small
\setlength{\tabcolsep}{2pt}
\begin{center}
\begin{tabular}{lcll}
batch & day & date & time $\pm$ range \\
\hline
5.0.5/5.0.6/5.5a5  & Tue & 22 Dec & 09:00 {\scriptsize UTC} / 17:00 {\scriptsize CST}  \\
5.5/6.0a1          & Fri & 29 Jan & 06:03 {\scriptsize UTC} / 14:03 {\scriptsize CST} $\pm$137\/s \\
5.5.1/5.5.2/6.0a2  & Tue & 23 Feb & 02:47 {\scriptsize UTC} / 10:47 {\scriptsize CST} $\pm$1\/s \\
5.5.4/6.0a4        & Tue & 29 Mar & 06:04 {\scriptsize UTC} / 14:04 {\scriptsize CST} $\pm$16\/s \\
5.5.5/6.0a5/6.0    & Wed & 01 Jun & 02:46 {\scriptsize UTC} / 10:46 {\scriptsize CST} $\pm$15\/s \\
6.0.5/6.5a3$^*$    & Thu & 22 Sep & 06:41 {\scriptsize UTC} / 14:41 {\scriptsize CST} $\pm$63\/s \\
6.0.6/6.5a4        & \multicolumn{3}{c}{\emph{various}} \\
6.0.8/6.5a6        & \multicolumn{3}{c}{\emph{various}} \\
\end{tabular}

\caption{
Common blocking times of bridges in a release batch.
Before October 2016, all bridges within a batch were blocked
within a few minutes of each other.
(The only exception is the 6.0.5/6.5a3 batch, marked with an `$^*$',
in which LeifEricson:50000 had been blocked since the beginning,
GreenBelt:5881 was offline during the time of the block,
and Azadi:4319 avoided notice.)
After that, an evident change in tactics caused
there to be no common date of blocking.
A shift to blocking by IP address meant that
some bridges were blocked even before having a ticket filed.
}
\label{table:batchBlockTime}
\end{center}
\end{table}

{}

\subsection{Bridge List File}
\label{readingbridgepref}

The bridge list file is a configuration file with all the bridges written on it.
Reading this file seems to be the easiest way for censors to discover new bridges.

In Section~\ref{sec:differenttimes},
we present evidence supporting that the GFW was not reading the bridge list file at that time.
It seems that it used black-box network analysis to find the bridges instead.

However, there has been recent evidence
to suggest that China operators have changed their behavior
and are now reading the this file
either instead of or in addition to black-box testing.
It appears as though they began looking at the bridge configuration file
between the release of 6.0.5/6.5a3 and the blocking of Azadi:6041.

By looking at Table~\ref{table:azadi},
we know they were not looking at the bridge list file before the release of 6.0.5/6.5a3
because some of the bridges in 6.0.5/6.5a3 did not get blocked (including Azadi:4319).
We already discussed this in section~\ref{sec:differenttimes}.

Starting from Tor 6.0.6/6.5a4, the censors started preemptively blocking bridges.
Azadi:6041 was released in version 6.0.6/6.5a4.
However, it got blocked after its ticket was merged and before the release came out.
Furthermore, the same ticket that added Azadi:6041
also removed Azadi:4319 from the configuration file.
Azadi:4319 was one of the bridges that did not get blocked in 6.0.5/6.5a3.
When Azadi:6041 was blocked, Azadi:4319 remained unblocked.
Even after 6.0.6/6.5a4 is released, Azadi:4319 was not blocked until
after Azadi was blocked on the whole IP address, on November 7, 2016.
The likely explanation is that they looked at the ticket or the Git repository for Tor.
Either way, this means that they were looking at the bridge list file to block new bridges.

Therefore, we conclude that
censors must have started looking at the bridge list file sometime between those
two events. This is a change from past behaviors and appears to be a new
action taken by China.

\begin{table}
\begin{center}
\begin{tabular}{lr}
\textbf{Event} & \textbf{Date} \\
\hline
Azadi:4319 opens & Aug 30 \\
6.0.5 public release (contains Azadi:4319) & Sep 16 \\
Rotate Azadi:4319$\rightarrow$Azadi:6041 & Oct 05 \\
Azadi:6041 blocked & Oct 20 \\
6.0.6 public release (contains Azadi:6041) & Nov 15 \\
\end{tabular}
\caption{
Azadi timeline
}
\label{table:azadi}

\end{center}
\end{table}

\subsection{Commented Bridges}

From Section~\ref{readingbridgepref}
we know that the censors changed their behavior
and started reading the bridge list file to find new bridges.
We wanted to know whether this process is automatic or manual.

We hypothesized the following:
\begin{itemize}
\item Censors parse the bridge list file automatically.
\item A person manually reads the bridge list file.
\end{itemize}

In the 6.0.6/6.5a4 release batch,
we incorporated two new bridges at the same time:
Lisbeth:443 and NX01:443.
We left NX01 commented out and added Lisbeth in as normal,
as seen below. 

\begin{verbatim}
pref(..., "obfs4 192.95.36.142:443 ...");
// Not used yet
// pref(..., "obfs4 85.17.30.79:443 ...");
\end{verbatim}

We discovered that of the two,
only Lisbeth was blocked while NX01 remained unblocked for a period of time.
NX01 was later blocked after it was uncommented.

Our reasoning is that this would help us
distinguish between human inspection and automatic blocking.
If humans are processing the source code manually,
they are likely to block NX01 and Lisbeth together.
If the blocking process is automatic,
then NX01 would be left unblocked.  

Since only the uncommented bridge Lisbeth was blocked initially,
this suggests an automatic parser rather than a manual parser.
We believe that had a person been manually viewing the file,
they would have also blocked NX01
since they would have seen that it was in the file.

\subsection{Other Anomalies}
\label{sec:anomalies}

\begin{figure}
\includegraphics{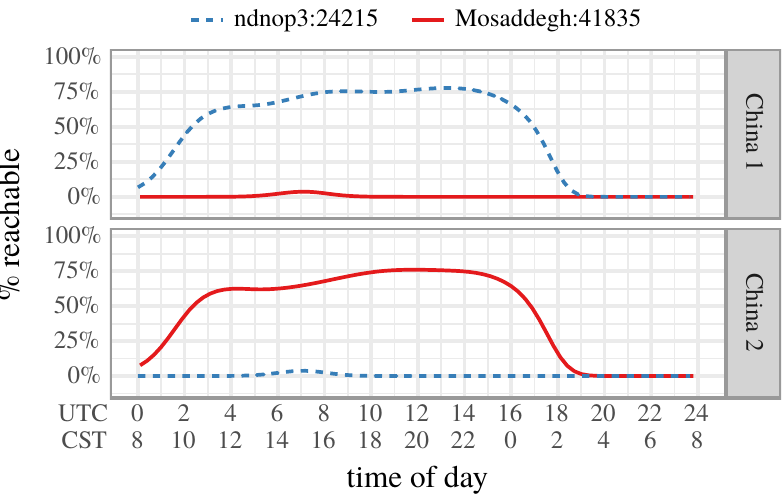}
\caption{
Rates of reachability by time of day
for two bridges from two sites,
between February~1 and March~15, 2016.
There is a diurnal blocking pattern
in both China sites,
though not the same bridges are affected at both sites.
China Standard Time (CST) is UTC+08:00.
}
\label{fig:diurnal}
\end{figure}

There is a conspicuous on--off pattern
in the reachability of certain bridges from China,
for example ndnop3:24215 between January~15
and May~10, 2016.
The pattern is roughly periodic with a period of 24~hours.
Figure~\ref{fig:diurnal} averages many 24-hour periods
to show the reachability against time of day of two bridges.
The presence of the diurnal pattern appears to depend on both the bridge and the probing site,
perhaps depending on the network path,
as the same bridges do not show the pattern at both sites.
The pattern can come and go, for example in riemann:443
before and after April~1, 2016.

The China sites also display what are apparently
temporary failures of censorship, stretches of a few hours
during which otherwise blocked bridges were reachable.
Intriguingly, one of these corresponds to a known
failure of the Great Firewall that was documented in the press~\cite{scmp-gfw}.
On March~27, Google services---usually blocked in China---were reachable from
about 15:30 to 17:15 UTC.
This time period is a subset of one in which our bridges
were reachable,
which went from about 10:00 to 18:00 UTC on that day.

\subsection{Censorship Measurement Platforms}
\label{sec:ooni-iclab}

The Open Observatory of Network Interference (OONI) is a project that
detects censorship around the globe. We contacted the the OONI team and they agreed
to help us with our experiment.
This afforded us more than 100 observation sites in different
countries that measured Tor bridge connectivity. This measurement
started in early December 2016 and we currently have measurement data for approximately three months. 

One really interesting observation we have is that
the firewalls of several countries might be faking connection responses.
These countries includes Thailand, Indonesia, Netherlands and Bulgaria.
This behavior is especially obvious in the probe data for the bridge fdctorbridge01:80.
The bridge fdctorbridge01:80, at that time, was defunct,
having closed down in May 2016~\cite{torbug-18976}.
Naturally we should get a connection error (timeout) when we try to probe it.
However, measurements in these few countries showed that the connection to port 80 at this bridge was successful.

Since we only had very few probe locations in Thailand, Indonesia, Netherlands,
it might be just a local HTTP proxy problem for our probe sites.
However, there are three probe sites in different ASes in Bulgaria,
and we noticed that this problem exists in all three probe locations.
This seems to indicate this isn't just a local problem for a specific network.
We decided to dig further into Bulgaria by bringing in a VPN,
and attempting to make TCP connections to fdctorbridge01.
We were not able to reproduce the results on OONI report, however,
since all of our connections timed out,
which is the expected behavior.

We also noticed that for the probe sites in Bulgaria (AS~44901),
connections to port 80 and 443 for certain IP addresses have an extremely short response time,
while connections to other ports takes much longer.
For example, we saw that connections to port 80 and 443 for MaBishomarim only takes around one millisecond.
Connections to other ports take around 500 milliseconds instead.
This substantial difference makes us wonder whether a HTTP spoofing is taking place.

\subsection{Retrospective Analysis}
\label{sec:retrospective}

\begin{center}
\begin{figure*}
\includegraphics[scale=0.4]{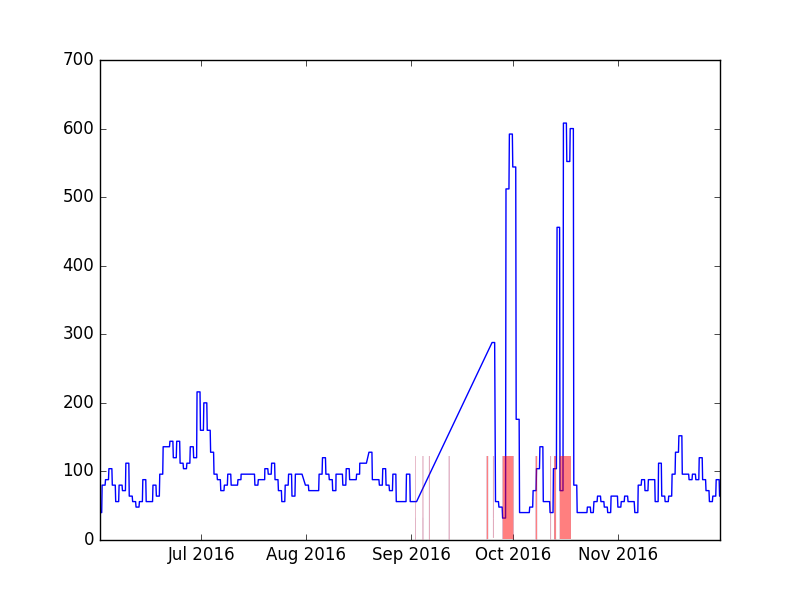}
\includegraphics[scale=0.4]{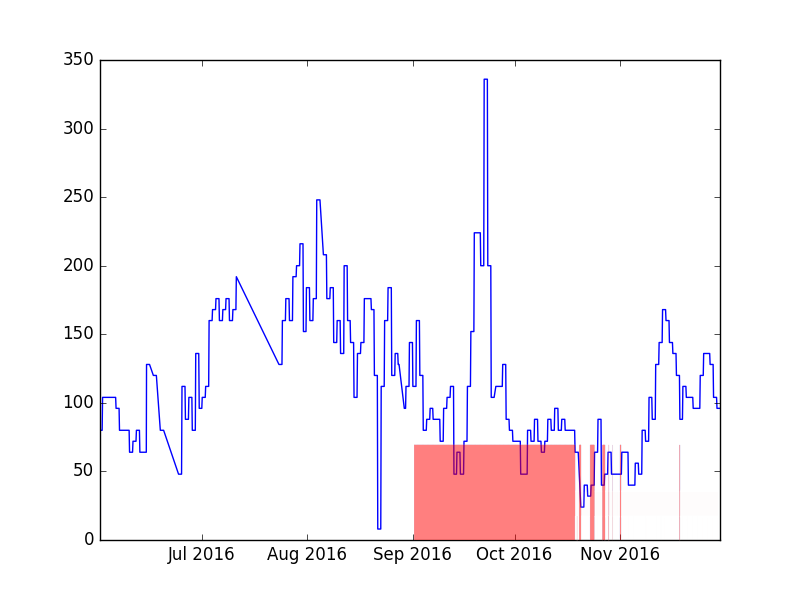}
\caption{
The left
graph is the daily traffic data for LeifEricson. The right one is for
MaBishomarim. The blue line is the number of daily connections from China.
The red bar is the actual connectivity data we measured. The long sloped
section from September to October for the traffic data in LeifEricson is an
artifact of missing data. This is no measurement data from CollecTor available
for that time and the line graph just connected the two nearest data points.
}
\label{fig:retrospective}
\end{figure*}
\end{center}

There is a gap in our measurement data from China during August 2016.
In order to
see the bigger picture and capture a larger timeframe, it is preferable that we
can look at logs of data even before December 2015. 

CollecTor~\cite{collector} is a public Tor data collection service.
It collects network status of Tor across the entire network.
It records the amount of daily traffic of each bridge.
More precisely,
it records the number of daily connections each bridge is
receiving from each country.
We believe that by looking at the number of daily connections from China,
we can estimate whether a bridge was blocked.
Since it is known that China does not have the same blocking rules across different areas~\cite{Ensafi2015a},
this data would contain a substantial amount of noise.
Furthermore, not all bridges have data for every single day.
Blocking inference from this would be a rough estimate at best,
but we still think that it would be useful to see the whole picture.

We used China's daily traffic data from CollecTor to infer the reachability of
different bridges. This can be supplementary to our own measurements. CollecTor
can be used to infer historical data that we were not able to see directly. The
logic behind this is that when a bridge is no longer reachable from China,
there would be a large drop in its daily traffic from China. Due to the fact
that China has different blocking rules in different regions, the amount of
traffic might never reach zero. However, we still think that the drop in
traffic should be large enough for us to distinguish between when the bridge is
blocked from China and when it is not.

From Figure 2, we can see CollecTor data and our own measurement for
LeifEricson. There is a clear match between the two. When our
measurement shows that the bridge is reachable, there would be a spike in the
traffic data. Looking at the CollecTor data, we see that the last time
LeifEricson was reachable from China was around July 2016. We know that LeifEricson 
was blocked on the whole IP even before our
monitoring started in September, but were unable to observe the time.
Looking at the CollecTor traffic data, we suspect the IP block may have
started much earlier than we expected, 
potentially as early as July 2016. 
%
%
%

Traffic data for MaBishomarim demonstrates the rare
case that the traffic data does not match our observation. From September to
October 2016, they still match well, but since November 2016, they
became contradictory. Our own measurement shows that MaBishomarim
has not been reachable since November 2016, but CollecTor shows that
MaBishomarim received a large burst of traffic in mid-November 2016. We
speculate that this incongruity is the artifact of the different blocking rules
in different regions of China. While the region our measurement site is in
blocked MaBishomarim, a different region still had MaBishomarim unblocked.
%
%
%
%
%
%

\section{Limitations}
\label{sec:limitations}

In our tests, we assume that a bridge is unblocked if we are able to make a
TCP connection. It may be the case, however, that a censor effectively
blocks a bridge despite allowing TCP connections. The fact that one can
connect to a bridge does not always mean that the censor will allow a
sustained obfs4 connection. We know of exactly such a case in
Kazakhstan where TCP reachability tests underestimate the level of
censorship. The firewall in Kazakhstan blocks Tor Browser's default obfs4
bridges, but differently from the GFW: it stops transferring
packets only some time after a connection is established~\cite{torbug-20348}.

We rented a VPN with an endpoint in Kazakhstan (AS~203087)
and ran tests from
December 18, 2016 to February 4, 2017. There is a risk that the censorship
seen by a VPN censorship may not be representative of censorship elsewhere in a
country; we have not been able to eliminate that possibility but we verified
that the VPN saw censorship of at least some domains like tumblr.com. We ran
our usual active probing experiments of all bridges every 20 minutes, as well
as hourly attempts to establish a Tor connection to a selection of
public and non-public bridges.
The results of TCP reachability tests were uniform:
all bridges were always reachable.
The results of the Tor bootstrap tests told a different and ambiguous story.
Tor measures its bootstrapping progress as a fraction between 0~and 100\%.
Some bridges, like Lisbeth:443 and Mosaddegh:9332, were essentially
always able to reach 100\%.
Others, like Mosaddegh:80, GreenBelt:80, and GreenBelt:5881, always
stalled at 10\% (indicating a failure after the initial TLS handshake).
Yet others, like Mosaddegh:443 and Mosaddegh:1984,
reached 25\%, showing that at least some Tor protocol data
flowed through the obfs4 channel,
but not enough to establish a full Tor circuit.



\section{Implications for Circumvention}

There are a number of implications of our findings
for the practice of censorship circumvention.
One is that even naive approaches
like Tor Browser's default bridges are effective against
many censors.

Another implication comes from observation of
how the Great Firewall discovers and blocks default bridges.
Whether it is done through black-box testing,
or inspection of source code, there are ways around,
involving complications on the client side.
For example, rather than strictly obeying the static
list of bridge addresses built in,
the client software could deterministically
compute some function of the bridge list,
perhaps varying over time,
whose output is the real set of addresses to use.
Suppose the bridge addresses automatically changed every day:
a censor doing a black-box test would also have to test
every day, else it would miss the new set of bridges.
Or suppose the censor parses a file containing a list of bridges:
a countermeasure is to tweak addresses 
so that they do not accurately reflect the addresses the client
goes to; adding 1 to each octet of the IP address, for example.
That would succeed until the censor devotes energy
to reverse engineer the tweaking algorithm.


Of course these simple, incremental countermeasures
are essentially security through obscurity,
only perpetuating the lamented cat-and-mouse game of censorship
and circumvention.
Is it even worth pursuing such strategies,
rather than looking at new circumvention techniques
that are hard for a censor to block in principle?
One answer is that yes,
as long as censors remain relatively slow and stupid,
and a little bit of investment of effort brings
a large amount of effective circumvention,
it is worth keeping at least a little bit ahead
of the censors, even if it means tweaking
conceptually broken systems.
Of course, we should set our sights farther,
and not allow such pursuits to fully distract us
from working on fundamental advances.
Ultimately, it may be worth it
to play the cat-and-mouse game
because by doing so and paying careful attention,
we learn surprising facts about censors
and their operation,
revealing weaknesses that can help the development of future systems.

\section{Ethics and Safety}

There are risks involved in running Tor
experiments. Some risks include disrupting other measurements, disclosing
bridge locations, and endangering the bridge operators.
During our experiment, we consulted with the
Tor Research Safety Board~\cite{torsafetyboard},
which helps researchers conduct experiments safely.
The research summary we sent to the board
is included in Appendix~\ref{sec:safety}.





{
\footnotesize \bibliographystyle{acm}
\bibliography{detecting-censor-detection}
}

\appendix
\section{Tor Research Safety Board}
\label{sec:safety}

This appendix contains a copy of the research summary we sent to the Tor
Research Safety Board~\cite{torsafetyboard}, a group of researchers who provide
recommendations on conducting research on Tor in a safe way.  It is included
here with no changes except for formatting.


\bigskip

We're seeking comments on a continuation of our research on the blocking
of default Tor Browser bridges. What we've done so far on this subject
is covered in our FOCI 2016 paper, ``Censors' Delay in Blocking
Circumvention Proxies'':
\url{https://www.bamsoftware.com/proxy-probe/}

The short summary of what we want to do is to greatly expand our
measurement locations, by using existing platforms such as ICLab, OONI,
or RIPE Atlas. We want to start doing traceroutes in addition to TCP
reachability. We want to control how new bridges are introduced, in
order to test specific hypotheses, such as whether there is a difference
in detection between stable and alpha.

\bigskip
\noindent
\textit{1.~What are you trying to learn, and why is that useful for the world? That is, what are the hoped-for benefits of your experiment?}

\begin{enumerate}
\item Where the default bridges are blocked, globally. We know that China
 (eventually) blocks them, and Iran (currently) does not; but we don't
 know the situation anywhere else.
\item In places where the default bridges get blocked, the dynamics of
 blocking, such as how long it takes, its granularity (IP only or
 IP/port), and whether blocks are eventually removed.
\item How bridge addresses are discovered (e.g. through traffic analysis,
 tickets, or source code), and how they are extracted (e.g. manually
 or through automated parsing).
 \end{enumerate}

The overarching, abstract benefit of the experiment is a better
understanding of censorship, leading to the development of better
informed circumvention.

The latest bridge users' guide
(\url{https://blog.torproject.org/blog/breaking-through-censorship-barriers-even-when-tor-blocked})
recommends using meek to users in China, because obfs4 is blocked. This
research would let us know whether to expand that advice beyond China.

By comparing reachability timelines across many censors, we may find
evidence for or against censors sharing a common data source. For
example, if two countries block a set of bridges at the same moment, it
is probably because there is something in common in their detection.

We may uncover specific operational weaknesses of censors that can be
exploited. To choose an invented but plausible scenario, maybe a censor
only does black-box testing of new bundles on the day of release: in
that case, the browser could avoid connecting to a subset of bridges
until after a certain date.

If we are able to reachability publish data online on a frequently
updated basis, someone could use it to build a Weather-like service that
notifies operators of default bridges when their bridge stops running.
This happened a few times already: some of the default bridges stopped
running because of lost iptables rules after a reboot, and we were the
first to notice, only because we were looking at the graphs every once
in a while. (This would not always be possible using only Collector
data, because for example the bridge might be running, but its obfs4
port closed because of a firewall misconfiguration.)

\bigskip
\noindent
\textit{2.~What exactly is your plan? That is, what are the steps of your experiment, what will you collect, how will you keep it safe, and soon.}

So far, we have only run from a handful of VPSes, never more than 4 at a
time. We only had visibility into the U.S., China, and Iran. We
carefully watched for the introduction of new obfs4 bridges (in some
cases being privately informed in advance), and added them to a probe
list, which got probed every 20 minutes by a cron job on the VPSes.

We want to greatly expand our probe sites, by using existing measurement
platforms such as ICLab, OONI, or RIPE Atlas. We hope to be able to
measure from dozens or hundreds of diverse locations. We have already
talked to ICLab and they are willing to probe our destinations from
their endpoints, which mostly consist of commercial VPNs in various
countries. The probes will consist of periodic TCP connections to Tor
Browser default obfs4 bridges (released and not-yet-released) and
control destinations. We want to start doing traceroutes as well.

We expect that the TCP reachability data we collect will be similar to
what we have collected so far. It looks like this:
{
\scriptsize
\begin{verbatim}
date,site,host,port,elapsed,success,errno,errmsg
1449892115.2,bauxite,178.209.52.110,443,10.0101830959,False,None,timed out
1449901202.36,eecs-login,192.30.252.130,443,0.0761489868164,True,,
1450858800.18,eecs-login,109.105.109.165,24215,0.189998865128,False,146,[Errno 146] Connection refused
\end{verbatim}
}
For traceroutes we will collect hop information (perhaps with some hops
obscured; see the risks in the next section). We expect to be able to
publish everything we collect in an immediate and ongoing basis.

We also want to test some specific hypotheses by controlling the
circumstances of bridge release. Here are specific experiments we have
thought of (see corresponding risks in the next section):
\begin{enumerate}
\item[a.] Rotating bridge ports with every release. Since the GFW blocks based
on IP/port, we can try just changing the port number of each bridge
in every release (using iptables forwarding for example).
\item[b.] Putting different subsets of bridges in stable and alpha releases. We
saw that Orbot-only bridges did not get blocked; we wonder if
stable-only or alpha-only bridges also will not get blocked.
\item[c.] Leaving a bridge commented out in bridge\_prefs.js. This may help us
distinguish between black-box testing and manual source code review.
\end{enumerate}

\bigskip
\noindent
\textit{3.~What attacks or risks might be introduced or assisted because of your actions or your data sets, and how well do you resolve each of them?}

The main risk is potentially enabling censors to discover new bridge
addresses early, by monitoring our probe sites. Even though ``default
bridges'' are conceptually broken, they do in fact work for many people,
and we wouldn't want to reduce their utility.

In our research so far, we've identified a number of ways that censors
can discover new bridges: by watching the bug tracker, by reading source
code, or by inspecting releases. Whenever possible, we want to start
monitoring new bridges even before they enter the bug tracker. If a
censor discovers one of our probe sites (which would not be hard to do),
then they could watch for new addresses being connected to and add them
to a blocklist. An adversary keeping netflow records could identify
probe sites retroactively: download Tor Browser and get the new bridges,
then find the clients that made the earliest connections to those
addresses.

We mitigate this risk partially by only testing default bridges, not
secret BridgeDB bridges. That way, even if a censor discovers them, it
doesn't affect users of secret bridges. Also, we suspect that, because
default bridges are, in theory, easily discoverable, adding another
potential discovery mechanism of medium difficult does not greatly
increase the risk of their being blocked.

If early blocking of bridges as a result of our experiment becomes a
problem, we can adjust the protocol, for example not to monitor bridges
in advance of their ticket being filed.

Our heretofore published data do not include the IP addresses of the
probe locations. The people who supplied us with the probe locations
asked us not to reveal them. Traceroute will make it harder to conceal
the source of probes in our published data. We can, for example, omit
the first few hops in each trace, but we don't know the best practices
along these lines. The potential harm to probe site operators is
probably less when we use existing measurement platforms rather than
VPSes acquired through personal contacts.

Our results may be contaminated by other experiments being run from the
same source address. The measurement platforms we propose to use already
are running various other experiments, so they may be treated
differently by firewalls. The most likely wrong outcome is that we
falsely detect a bridge being blocked, when it is really the client
address being blocked (because it is a VPN node, for example). The risk
goes in the other direction as well: our experiment might affect others
running on the same endpoint.

Here are the risks related to testing the specific bridge-blocking
hypotheses enumerated in the previous section:
\begin{enumerate}
\item[a.] The risk in rotating bridge ports is that eventually the censor
catches on to the pattern and develops more sophisticated, automated
blocking. If the censor doesn't react, it means we have better
reachability; but if it does, we lose what small window of
post-release reachability we have.
\item[b.] The risk in segregating bridge addresses across stable and alpha is
that a network observer can tell which a user is running by observing
what addresses they connect to. This may, for example, enable them to
target an exploit that only works on a specific version.
\item[c.] The risk in playing games like commenting out bridge lines is slight:
a commented-out bridge may get blocked even before it has had any
real users.
\end{enumerate}

\bigskip
\noindent
\textit{4.~Walk us through why the benefits from item 1 outweigh the remaining risks from item 3: why is this plan worthwhile despite the remaining risks?}

The main risk, bridge discovery by censors, has low potential harm, and
can be mitigated if necessary by changing when we start monitoring
bridges, or even ceasing the experiment altogether. The risk of our
measurements is probably less than that of even having default bridges
in the first place, because our probes are not connected to any
real-world circumventor.

The risks associated with our specific bridge-blocking hypotheses are
variable, and we would appreciate discussion on them. The one we planned
to try first is the commenting-out one, because it seems to have the
best risk/reward tradeoff.

Incidentally, OONI already has a bridge\_reachability nettest that is
similar to what we have proposed: \url{https://ooni.torproject.org/nettest/tor-bridge-reachability/} However their bridge list is not up to date, \url{https://gitweb.torproject.org/ooni-probe.git/tree/var/example_inputs/bridges.txt?id=v1.6.1} and a perusal of \url{http://measurements.ooni.torproject.org/} shows that the test is not being run regularly.

\section{Code and Data}

\url{https://www.bamsoftware.com/proxy-probe/}

\end{document}